\documentclass[pre,11pt,twocolumn,reprint]{revtex4-1}

\usepackage[a4paper]{geometry}
\usepackage{amsmath}
\usepackage{amssymb}
\usepackage{amsbsy}
\usepackage{amsbsy}
\usepackage{mathrsfs}
\usepackage{multirow}
\usepackage{bm}
\usepackage{tikz}
\usetikzlibrary{matrix}
\usetikzlibrary{shapes,arrows}
\usepackage{lmodern}
\usepackage[english]{babel}
\usepackage{graphicx}
\usepackage{xcolor}
\usepackage{enumerate}
\tikzset{ 
	table/.style={
		matrix of nodes,
		row sep=-\pgflinewidth,
		column sep=-\pgflinewidth,
		nodes={rectangle,text width=3em,align=center},
		text depth=1.25ex,
		text height=2.5ex,
		nodes in empty cells
	},
	row 1/.style={nodes={fill=green!10,text depth=0.4ex,text height=2ex}},
	row 2/.style={nodes={text depth=2.4ex,text height=5ex}},
	row 3/.style={nodes={text depth=2.4ex,text height=5ex}},
	row 4/.style={nodes={text depth=2.4ex,text height=5ex}},
	row 5/.style={nodes={text depth=2.4ex,text height=5ex}},
	row 6/.style={nodes={text depth=2.4ex,text height=5ex}},
}

\begin{document}

\title{Virial-like Thermodynamic Uncertainty Relation in the Tight-Binding Regime}
\author{N. J. L\'opez-Alamilla$^{\rm\, a}$}
\author{R. U. L. Cachi$^{\rm\, a,b}$}
\affiliation{$^{\rm\, a\,}$Department of Physics, University of Otago, P. O. Box 56, Dunedin 9054, New Zealand}
\affiliation{$^{\rm\, b\,}$Department of Chemistry, KU Leuven, Leuven, Belgium}

\begin{abstract}
We presented a methodology to approximate the entropy production for Brownian motion in a tilted periodic potential.
The approximation stems from the well known thermodynamic uncertainty relation. By applying a virial-like expansion, we provided a tighter lower limit solely in terms of the drift velocity and diffusion. The approach presented is systematically analysed in the tight-binding regime. We also provide a relative simple rule to validate using the tight-binding approach based on drift and diffusion relations rather than energy barriers and forces. We also discuss the implications of our results outside the tight-binding regime. 
 
\end{abstract}

\date{\today}	
\maketitle

\section{Introduction}

Recently, fluctuation theorems have attracted a great deal of research in non-equilibrium physics\,\cite{BaratoSeifert2015,BaratoSeifert2016,PietzonkaBaratoSeifert2016,PietzonkaSeifert2018,GingrichHorowitzPerunovEtAl2016,RosasVandenBroeckLindenberg2017,HwangHyeon2018,BrandnerHanazatoSaito2018,FischerPietzonkaSeifert2018}. In particular, the \emph{ thermodynamic uncertainty relations} (TURs) that provide a trade-off between the cost (entropy generation) and the precision of the system\,\cite{BaratoSeifert2015,BaratoSeifert2016,PietzonkaBaratoSeifert2016,PietzonkaSeifert2018}. In the original derivation of the TUR the system considered was a motor protein,\,\cite{BaratoSeifert2015,GingrichHorowitzPerunovEtAl2016} and it has been shown that the TUR provides a lower bound for the entropy generation of this system. Generalizations of the TUR have been proposed  recently, for systems with multiple degrees of freedom,\,\cite{Dechant2018} quantum systems,\,\cite{MacieszczakBrandnerGarrahan2018,FalascoEspositoDelvenne2020,LeeHaJeong2021} and two bodies interacting systems\,\cite{SaryalSadekarAgarwalla2021}. However, experiments  and analytic studies show that systems usually operate far above the lower bound predicted by the TUR\,\cite{SongHyeon2020,JackLopez-AlamillaChallis2020}, additionally, violations to this bound are possible in quantum systems\,\cite{PaneruDuttaTlustyEtAl2020,CangemiCataudellaBenentiEtAl2020,Kalaee2021}. 

For the case of motor proteins, we can consider them as overdamped Brownian particles  converting chemical energy into work\,\cite{SvobodaBlock1994,Magnasco1994,Astumian1998,ItohTakahashiAdachiEtAl2004,ToyabeEtAl2011,Kolomeisky2013}. Usually, the motor protein  will transform the released energy from the hydrolysis of fuel molecules (ATP or GTP) into the mechanical motion of the motor by a specific distance `$\delta$' commonly refereed as step size ($\delta\sim8\,{\text{nm}}$ for kinesin/dynein and $50\sim90\,{\text{nm}}$ for myosin, $\delta\sim2\pi/3\,{\text{rad}}$ for F$_0$F$_1$ATPase). When this chemical energy conversion is tightly-coupled, the motion of the motor can be effectively described as one-dimensional\,\cite{Magnasco1994}. In the long time limit, the system will reach  a steady-state, which in turn can be described by its rate of entropy production, drift velocity and diffusion\,\cite{ChallisJack2013,NguyenChallisJack2016,Challis2018}. 

Several descriptions of the directed Brownian motion observed in motor proteins have been proposed\,\cite{Astumian1998,Astumian2007,AstumianMukherjeeWarshel2016}. One widely spread approach is to  consider over-damped Brownian motion over a time-independent tilted periodic free-energy potential. From this free-energy approach one can numerically evaluate or in some cases derive closed solutions for steady-state dynamics\,\cite{ReimannVandenBroeckLinkeEtAl2001}. In this paper, we will use this approach in a system for which exact solutions exist. Next, we will briefly review the main issues surrounding the usage of the TUR entropy production lower bound. Then, we will introduce a formalism to derive a tighter  bound based on a virial-like expansion of the original TUR and its application.     

\section{background}

We are interested in the steady-state rate of entropy generation `$\sigma^{\rm ss}$' for over-damped Brownian motion over a tilted periodic potential 
\begin{equation}\label{Eq:V:PerTilt}
V(\hat x)=V_0(\hat x)-k_{\rm B}T\hat f\hat x,
\end{equation} 
where $\hat x=x/\delta$ is the dimensionless position coordinate, and  $\hat f=f\delta/k_{\rm B}T$ the dimensionless force, driving the system out of equilibrium, $V_0(\hat x+\delta)=V_0(\hat x)$ is the periodic part, $k_{\rm B}$ is the Boltzmann constant and $T$ the temperature. For a free-energy potential of the form of Eq.\,(\ref{Eq:V:PerTilt}), it has been shown that the rate of entropy generation has the form\,\cite{Gardiner2009,BaratoSeifert2015}    
\begin{equation}\label{Eq:S:gen}
\sigma^{\rm ss}=k_{\rm B}\hat v\hat f,
\end{equation}
with $\hat{v}=v/\delta$ the steady-state drift velocity. For the same system, the TUR cost-precision trade-off `$\mho$' is stated as follows\,\cite{BaratoSeifert2015,BaratoSeifert2016}
\begin{equation}\label{Eq:TUR:TradeOff}
\mho=\sigma^{\rm ss}\times\frac{\hat{D}}{\hat{v}^2}\geq k_{\rm B},
\end{equation}
with $\hat{D}=D/\delta^2$ the steady-state diffusion. 

From the above inequality, we obtain a lower bound `$\sigma^{\rm ss}_{\textsc{tur}}$' for the entropy generation of the system\,\cite{BaratoSeifert2015,BaratoSeifert2016} 
\begin{equation}\label{Eq:TUR:Bound}
\sigma^{\rm ss}\geq \sigma^{\rm ss}_{\textsc{tur}},
\end{equation} 
with 
\begin{equation}\label{Eq:S:TUR}
\sigma^{\rm ss}_{\textsc{tur}}=k_{\rm B}\frac{\hat{v}^2}{\hat{D}}.
\end{equation}
The usefulness of Eq.(\ref{Eq:TUR:Bound}) and its tightness for several regimes and systems has been explored\,\cite{JackLopez-AlamillaChallis2020}. Here, we will briefly review some important aspects of this bound.

Let us inspect the steady-state features of a model system. For simplicity, we will assume that Eq.\,(\ref{Eq:V:PerTilt}) periodic part $V_0(\hat{x})$ is: 
\begin{equation}\label{Eq:V:Eql}
	V_0(\hat{x})=k_{\rm B}TA_0\cos(2\pi(\hat{x}-\alpha)). 
\end{equation}
With such periodic part, the potential minima points are at $\hat{x}_{\rm c}=1/2+\arcsin{(\hat{f}/8\pi)}/2\pi\pm n$ and the maxima points are at $\hat{x}_{\rm a}=1-\arcsin{(\hat{f}/8\pi)}/2\pi\pm n$, with $n\in\mathbb{Z}$ (we have assumed $\delta=1$). Let us denote the energy barriers in the direction of the driving force of this potential as $E_{\rm a}(\hat{f}\,)=V(\hat{x}_{\rm a})-V(\hat{x}_{\rm c})$, in particular $E_{\rm a}(0)=2A_0$. In Figure 1 we show this free-energy potential. It is convenient to define the mean time for the system to cross  barrier $E_{\rm a}$ at equilibrium $\tau_0$ defined by
\begin{equation}
\tau_{0}=\frac{\gamma}{k_{\rm B}T}\int_{\hat{x}_{\rm c}}^{\hat{x}_{\rm a}}d\hat{y}\,e^{V_0(\hat{y})/k_{\rm B}T}\int_{\hat{y}}^{\hat{x}_{\rm a}}d\hat{z}\,e^{-V_0(\hat{z})/k_{\rm B}T}\,,
\end{equation}
with $\gamma$ the viscous drag coefficient, and the associated transition rate is $\kappa_0=\tau_0^{-1}$.
\begin{figure}[htbp]
	\begin{center}
		\includegraphics[width=0.45\textwidth]{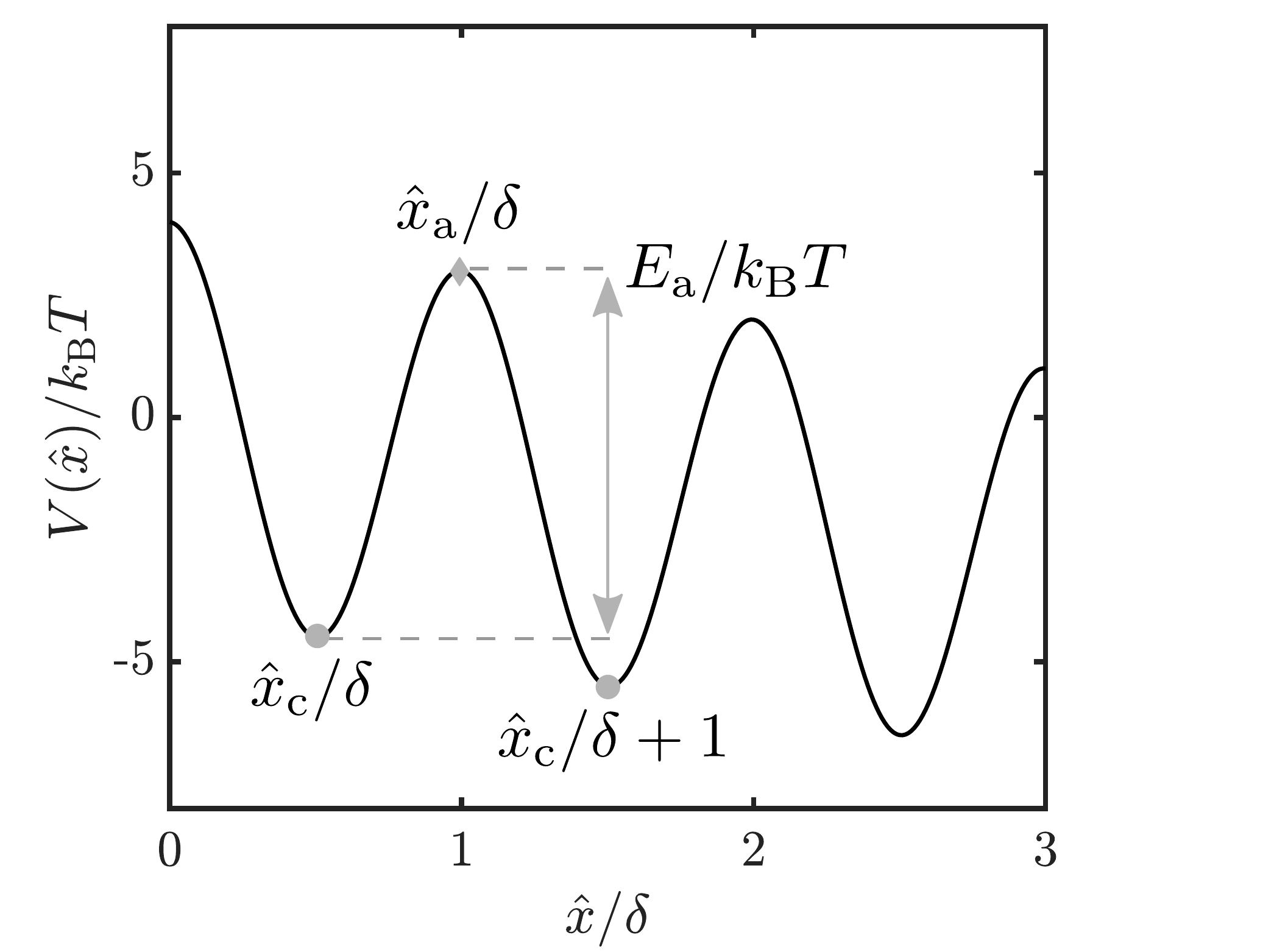}
		\caption{Scheme of energy potential Eq.\,(\ref{Eq:V:PerTilt}), with periodic part Eq.\,(\ref{Eq:V:Eql}), for reference we show critic points of the potential and energy barrier in the direction of the applied tilting force.}
		\label{Fig:1}
	\end{center}
\end{figure}

\begin{figure}[htbp]
		\begin{center}
			\includegraphics[width=0.45\textwidth]{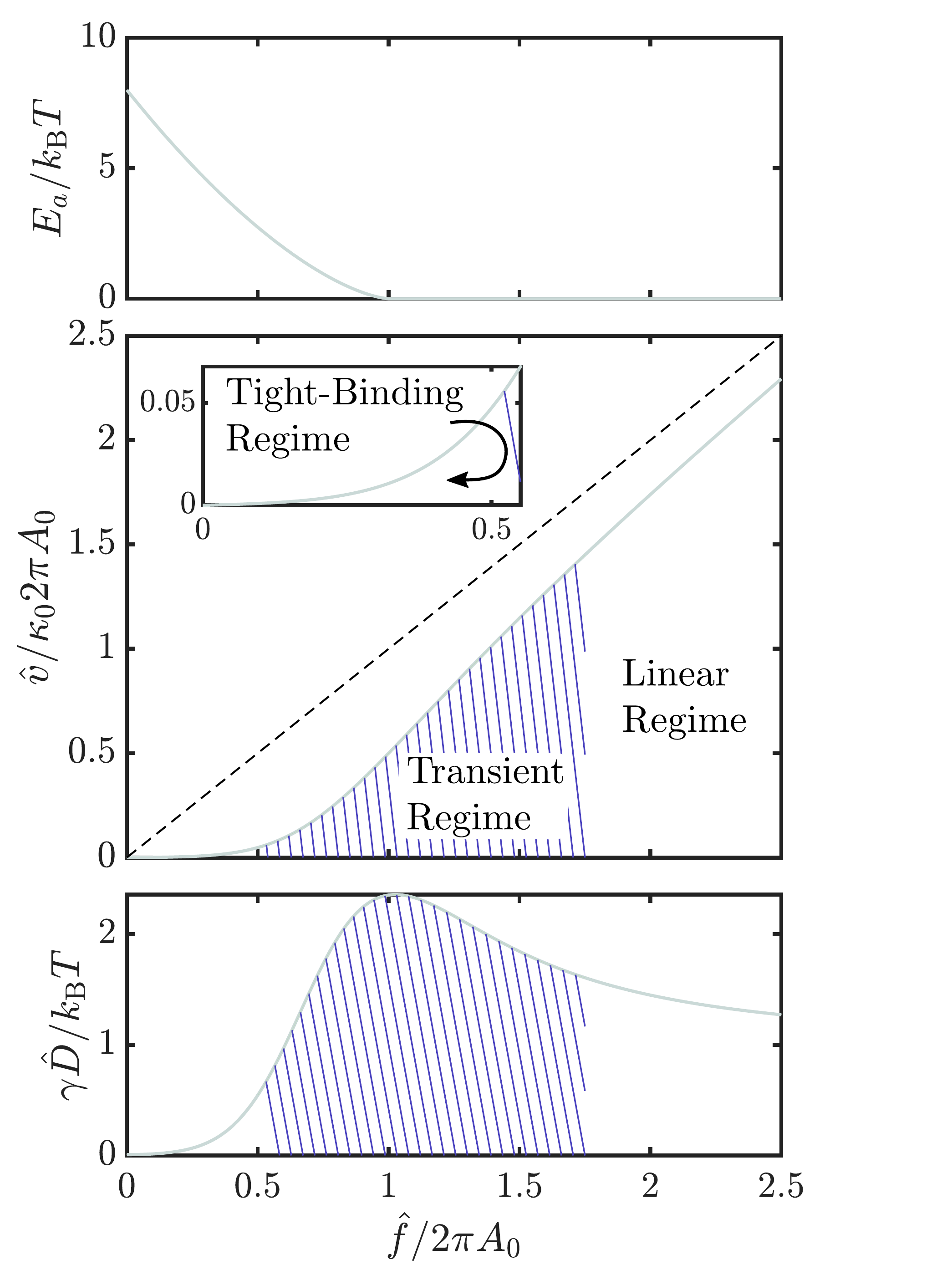}\put(-60,75){(c)}\put(-60,200){(b)}\put(-60,280){(a)}
			\caption{Characteristic properties of Brownian motion over an energy potential Eq.\,(\ref{Eq:V:PerTilt}) with periodic part Eq.\,(\ref{Eq:V:Eql})  as function of the tilting force. (a) Energy barrier. (b) Steady-state drift velocity, showing (patched) transient regime, linear regime and (inset) tight-binding regime, for reference (dashed) identity function for the tilting force. (c) Steady-state diffusion. Parameters used $A_0=4$.}
			\label{Fig:2}
		\end{center}
\end{figure}

By applying a variable force on potential Eq.\,(\ref{Eq:V:PerTilt}) and periodic part Eq.\,(\ref{Eq:V:Eql}), we can observe how the  energy barrier $E_{\rm a}$, drift velocity $\hat{v}$ and diffusion $\hat{D}$ evolve. This is shown in Figure~\ref{Fig:2}.
The values of $\hat{v}$, $\hat{D}$ where evaluated from the closed solutions by Stratonovich \,\cite{Stratonovich1958} and Reimann, et al \,\cite{ReimannVandenBroeckLinkeEtAl2001} respectively. We can identify three different regimes for the drift velocity. These regimes correspond to the particular features observed in each of them. In the tight-binding regime both drift and diffusion grow exponentially with the force\,\cite{Gardiner2009,Challis2016}. In the transient regime the diffusion display an stochastic resonance\,\cite{ReimannVandenBroeckLinkeEtAl2001}. Finally, in the linear regime $\hat{v}\propto\hat{f}$, and $\hat{D}\rightarrow k_{\rm B}T/\gamma$\,\cite{ReimannVandenBroeckLinkeEtAl2001}. 

\subsection{Entropy generation and the TUR}

\begin{figure}[htbp]
	\begin{center}
		\includegraphics[width=0.45\textwidth]{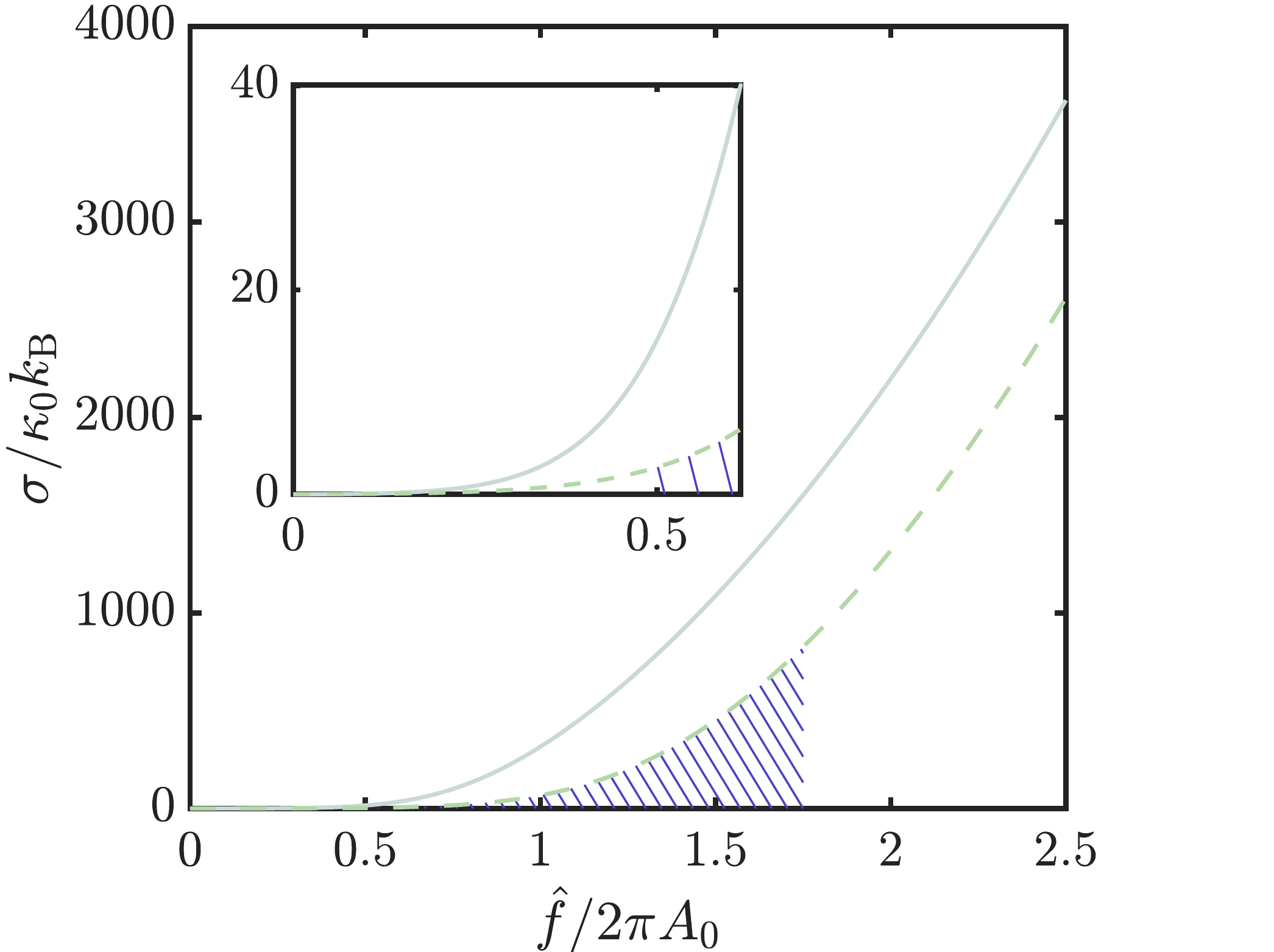}
		\caption{Steady-state entropy production of potential Eq.\,(\ref{Eq:V:PerTilt}) with periodic part Eq.\,(\ref{Eq:V:Eql}) as function of the tilting force, with (solid) Eq.\,(\ref{Eq:S:gen}) and (dashed) Eq.\,(\ref{Eq:S:TUR}). Inset shows the tight-binding regime. Parameters used $A_0=4$.}
		\label{Fig:3}
	\end{center}
\end{figure}

Now, let us compare Eqs.\,(\ref{Eq:S:gen}) and (\ref{Eq:S:TUR}) for variable $\hat{f}$, see Figure~\ref{Fig:3}. 
As we can see, even in the tight-binding regime, the lower bound provided by the TUR is a relative loose bound. One can expect that for very small force $\hat{f}\rightarrow0$, $\sigma^{\rm ss}\gtrsim\sigma^{\rm ss}_{\textsc{tur}}$, implying that $\hat f\gtrsim \hat{v}/\hat{D}$. 

Let us look at potential Eq.\,(\ref{Eq:V:PerTilt}) with periodic part Eq.\,(\ref{Eq:V:Eql}). By setting $A_0\gg1$ and $\hat {f}\approx0$, this corresponds to the tight-binding regime. Here, $\hat{v}$ and $\hat{D}$ have the following functional dependence on the driving force  
\begin{equation}\label{Eq:TB:drift}
\hat{v}=2\kappa_0\sinh\left(\frac{\hat f}{2}\right),
\end{equation}
\begin{equation}\label{Eq:TB:diffusion}
\hat{D}=\kappa_0\cosh\left(\frac{\hat f}{2}\right).
\end{equation}
In this regime  the ratio $\hat{v}/\hat{D}$ becomes
\begin{equation}
\frac{\hat{v}}{\hat{D}}=2\tanh\left(\frac{\hat f}{2}\right).
\end{equation}  
If we consider the Taylor series of $\tanh(\hat f/2)$ for small $\hat f$ values we have
\begin{equation}\label{Eq:Taylor}
2\tanh\left(\frac{\hat f}{2}\right)\approx \hat f-\frac{\hat f^3}{12}+\frac{\hat f^5}{120}-\ldots
\end{equation}
if $\sigma^{\rm ss}\gtrsim\sigma^{\rm ss}_{\textsc{tur}}$   Eq.\,(\ref{Eq:Taylor}) shows that agrees with $\hat{f}$ as a 1st order approximation. However, the other terms of the Taylor series are not insignificant, explaining  the increasing disagreement  between Eq.\,(\ref{Eq:S:gen}) and Eq.\,(\ref{Eq:S:TUR}) as $\hat{f}$ increases observed in Figure~\ref{Fig:3}. This opens the question, is there a comprehensive way to approximate $\hat{f}$ in terms solely of $\hat{v}$ and $\hat{D}$? 

\section{A tighter entropy bound using a Virial-like TUR approach}

\begin{figure}[htbp]
	\begin{center}
		\includegraphics[width=0.45\textwidth]{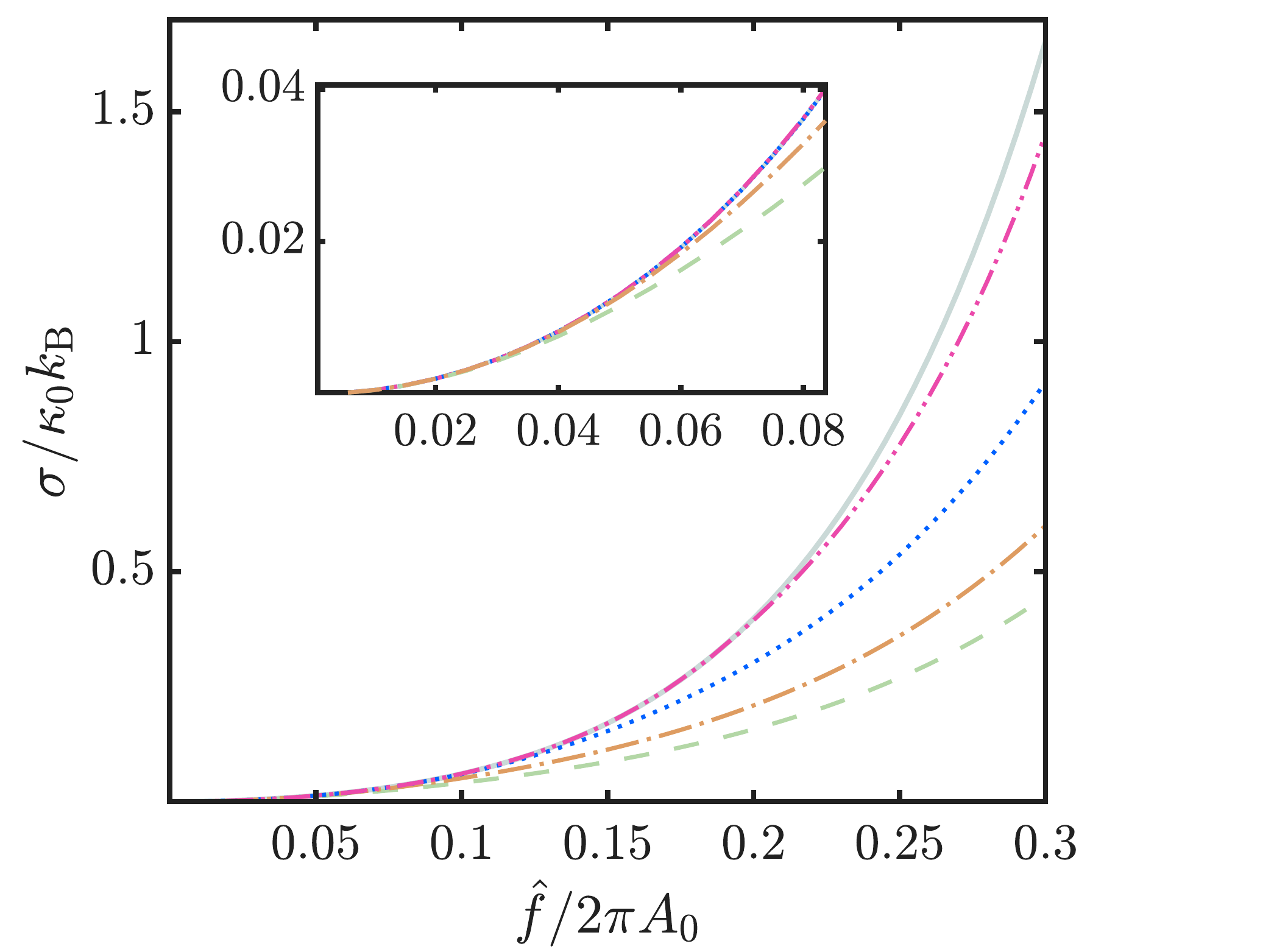}
		\caption{Steady-state entropy production of potential Eq.\,(\ref{Eq:V:PerTilt}) with periodic part Eq.\,(\ref{Eq:V:Eql}) as function of the tilting force, via: (solid) Eq.\,(\ref{Eq:S:gen}), (dashed) TUR Eq.\,(\ref{Eq:S:TUR}), (dash-dotted) Eq.\,(\ref{Eq:Virial:TUR}) 2-nd order, (dotted) Eq.\,(\ref{Eq:Virial:TUR}) 7-th order, (dash-dot-dot-dashed) Eq.\,(\ref{Eq:Virial:TUR}) 29-th order. Parameters used $A_0=4$.}
		\label{Fig:4}
	\end{center}
\end{figure}

Let us start from the assumption that we can express $f$ as a series of $g$ functions that depend on both $v$ and $D$.
\begin{equation}
\hat f\approx\sum_{n=1}^{N} g(a_nv^n,b_nD^n)+\mathcal{O}^{N+1}
\end{equation}
Looking at the Taylor series for $\tanh(f/2)$, it is clear that by raising to the right power $\tanh(f/2)$ and multiplying by a proper coefficient, one can reproduce the higher order terms in expansion Eq.\,(\ref{Eq:Taylor}). Thus, after some cumbersome algebra, we find that the following sum satisfies 
\begin{equation}\label{Eq:Virial:TUR:Krammers}
\frac{\hat f}{2}\gtrsim \tanh\left(\frac{\hat f}{2}\right)+\frac{\tanh\left(\frac{\hat f}{2}\right)^3}{3}+\frac{\tanh\left(\frac{\hat f}{2}\right)^5}{5}+\ldots
\end{equation} 
in other words,
\begin{equation}\label{Eq:Virial:Force}
\lim\limits_{\hat{f}\rightarrow0}\hat f\gtrsim\sum_{n=1}^{N} \frac{\hat{v}^{2n-1}}{2^{2n-2}(2n-1)\hat{D}^{2n-1}}+\mathcal{O}^{N+1}
\end{equation}
Then, in the regime where Eqs.\,(\ref{Eq:TB:drift})-(\ref{Eq:TB:diffusion}) are valid, we can provide a better approximation to the entropy generation of the system by implementing a higher and higher order sum of Eq.\,(\ref{Eq:Virial:Force}), resulting in a corrected lower bound of $N$-th order
\begin{equation}\label{Eq:Virial:TUR}
k_{\rm B}\hat{v}\hat f\geq\sigma^{N\text{-th}}_{\textsc{tur}}=k_{\rm B}\sum_{n=1}^{N} \frac{\hat{v}^{2n}}{2^{2n-2}(2n-1)\hat{D}^{2n-1}}
\end{equation} 
In Figure~\ref{Fig:4} we compare the bounds provided by Eq.\,(\ref{Eq:Virial:TUR}) with the original TUR bound Eq.\,(\ref{Eq:S:TUR}) and the actual entropy generation for the system Eq.\,(\ref{Eq:S:gen}) in the Tight-Binding regime. Notice that for $N=1$, Eq.\,(\ref{Eq:Virial:TUR})= Eq.\,(\ref{Eq:S:TUR}). 

\begin{figure}[htbp]
	\begin{center}
		\includegraphics[width=0.45\textwidth]{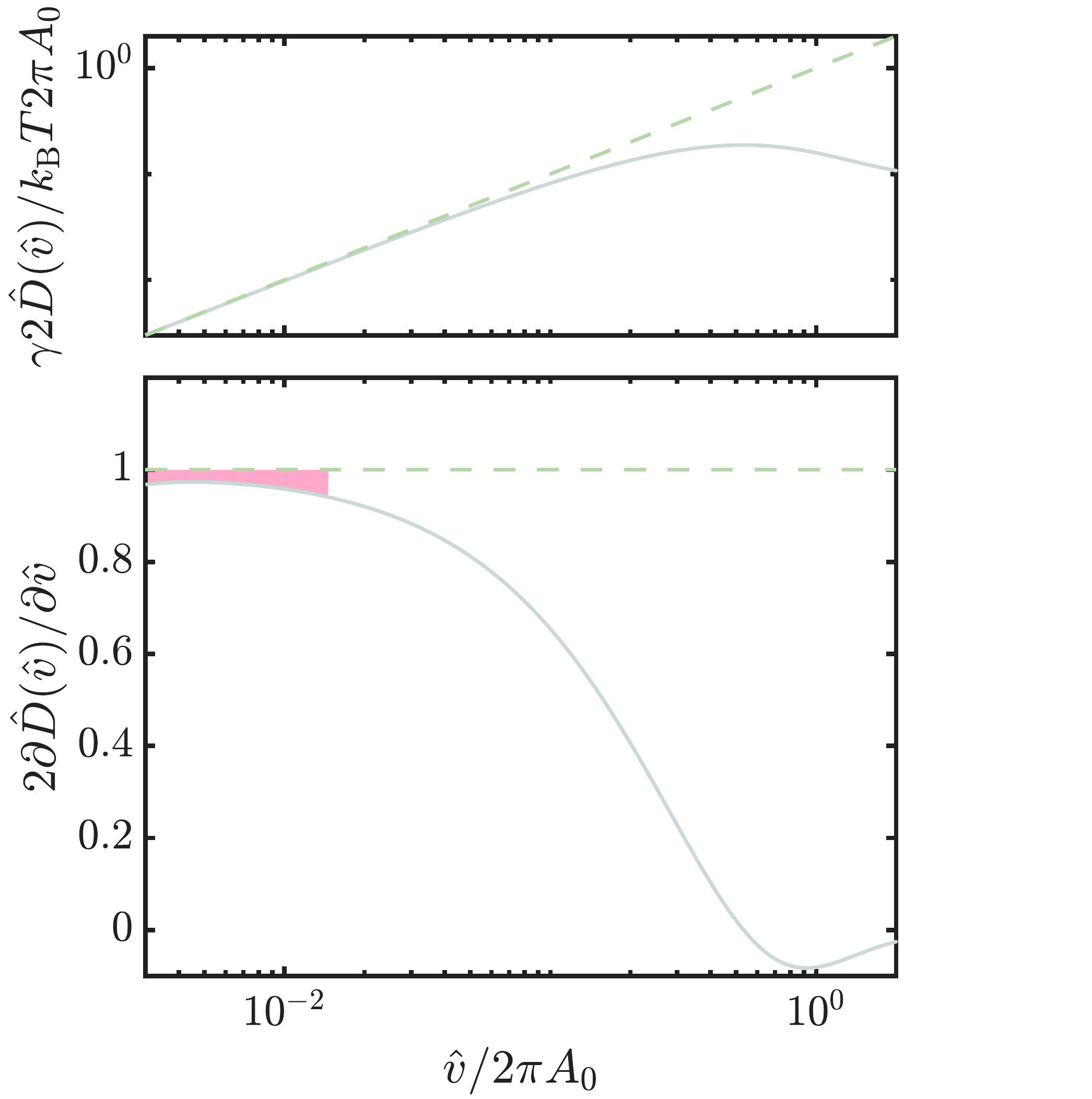}
		\caption{Evaluation of tight-binding regime validity for Eq.\,(\ref{Eq:V:PerTilt}) with periodic part Eq.\,(\ref{Eq:V:Eql}), showing: (a) Log-log plot of (solid) $2\hat{D}(\hat{v})$ and (dashed) $\hat{v}$. (b) Semi-log plot fo (solid) $2\partial\hat{D}(\hat{v})/\partial\hat{v}$ and (dashed) $\partial\hat{v}/\partial\hat{v}$ for reference. Shaded region shows $2\partial\hat{D}(\hat{v})/\partial\hat{v}\approx1\pm0.1$. Parameters used $A_0=4$.}
		\label{Fig:5}
	\end{center}
\end{figure}

One of the applications of the TUR lower bound is to be able to estimate the entropy production in the system without having knowledge of applied force value. Since, Eq.\,(\ref{Eq:Virial:TUR}) validity depends on being in the tight-binding regime. We need to understand how to evaluate if the system is in this regime without prior knowledge of the force value. The relative height of the energy barrier $E_{\rm a}$ (see Figure~\ref{Fig:1}) compared to the thermal energy has previously been used to validate the usage the tight-binding approach\,\cite{Challis2016,Challis2018}. Based on this metric, many authors set $E_{\rm a}\geq3k_{\rm B}T$  as a valid regime\,\cite{Astumian2007,AstumianMukherjeeWarshel2016,Challis2018}, this requires knowledge of the energy landscape which in many cases is unknown too. Here, we aim to systematically determine if we are in the tight-binding regime using only the observables of the system. Equation\,(\ref{Eq:Virial:TUR})  will only converge with increasing $N$-terms if $2\hat{D}$ and $\hat{v}$ grow at a similar rate. Thus, we will focus on the grow rate of $2\hat{D}$ as $\hat{v}$ changes, in other words $2\partial\hat{D}(\hat{v})/\partial\hat{v}$. Regardless of how the drift changes,   $\partial\hat{v}/\partial\hat{v}\equiv1$, then $2\hat{D}$ and $\hat{v}$ growing at a similar rate implies 
\begin{equation}\label{Eq:TB:test}
\frac{2\partial\hat{D}(\hat{v})}{\partial\hat{v}}\approx1\,,
\end{equation}
ensuring we are in the tight-binding regime. In Figure~\ref{Fig:5} we examine $d\hat{D}(\hat{v})/d\hat{v}$ for the model periodic potential Eq.\,(\ref{Eq:V:Eql}), we display this plot in a log scale  for readability purposes.  


\subsection{Beyond the tight-binding regime}

\begin{figure}[htbp]
	\begin{center}
		\includegraphics[width=0.45\textwidth]{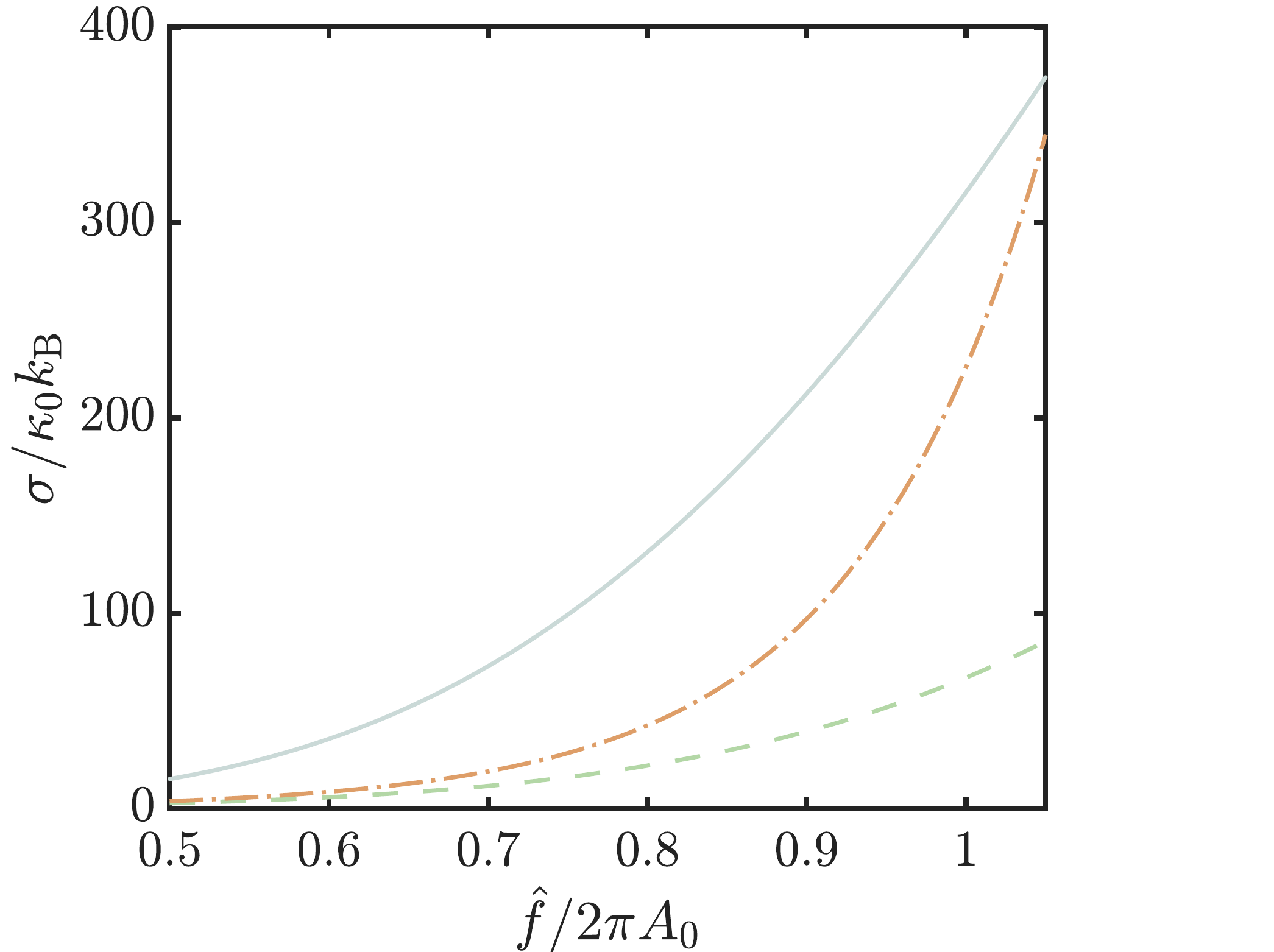}
		\caption{Steady-state entropy production potential Eq.\,(\ref{Eq:V:PerTilt}) with periodic part Eq.\,(\ref{Eq:V:Eql}) in the transient regime, via: (solid) Eq.\,(\ref{Eq:S:gen}) ,(dashed) TUR Eq.\,(\ref{Eq:S:TUR}), (dash-dotted) Eq.\,(\ref{Eq:Virial:TUR}) 2-nd order. Parameters used $A_0=4$.}
		\label{Fig:6}
	\end{center}
\end{figure}

Outside the tight-binding regime the approach presented in this paper starts to break down. In the linear regime $\hat{v}$ grows faster than $\hat{D}$ which additionally starts to decrease, see Figure~\ref{Fig:2}. Thus, it is evident that Eq.\,(\ref{Eq:Virial:Force}) and therefore Eq.\,(\ref{Eq:Virial:TUR}) will diverge as we consider larger $N$ values in the sum. In the transient regime if we evaluate Eq.\,(\ref{Eq:Virial:TUR}) up to very large $N$ values we encounter the same issue, however by cutting down the sum at 2-nd or 3-rd order improves the approximation compared to the TUR but ultimately diverges if the force is larger than the critic force $\hat{f}>2\pi A_0$. This is shown in Figure~\ref{Fig:6}. 

\subsection{Bi chromatic potential}

\begin{figure}[htbp]
	\begin{center}
		\includegraphics[width=0.45\textwidth]{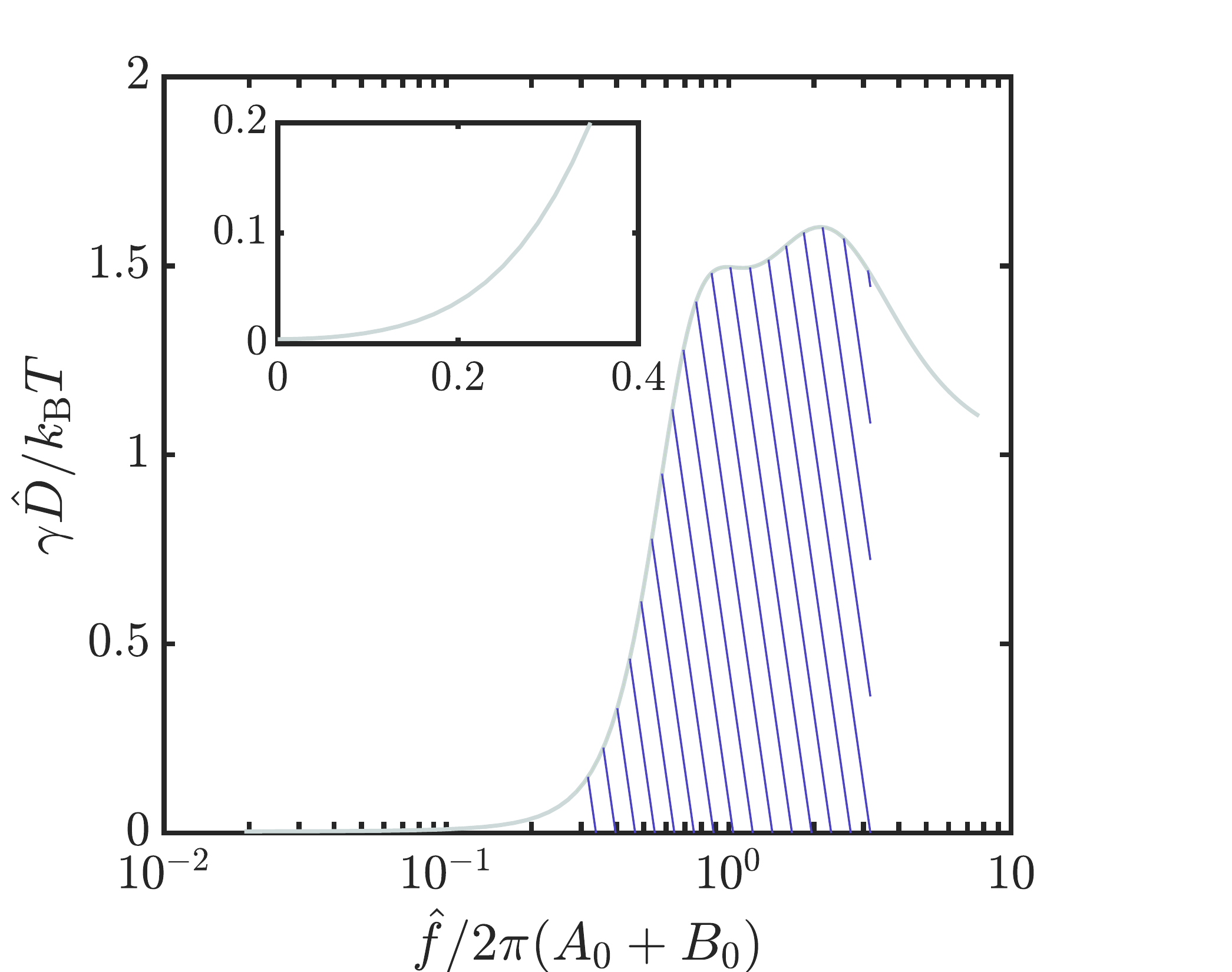}
		\caption{Diffusion as function of tilting force with for Eq.\,(\ref{Eq:V:PerTilt}) with periodic part Eq.\,(\ref{Eq:V:Bi}), showing: (patched) transient regime, (inset) the tight-binding regime and linear regime. The horizontal axis is in log$_{10}$ scale for readability purposes. Parameters used $A_0=4$, $\nu_A=1$, $B_0=1.2$, $\nu_B=8$.}
		\label{Fig:7}
	\end{center}
\end{figure}

\begin{figure}[htbp]
	\begin{center}
		\includegraphics[width=0.45\textwidth]{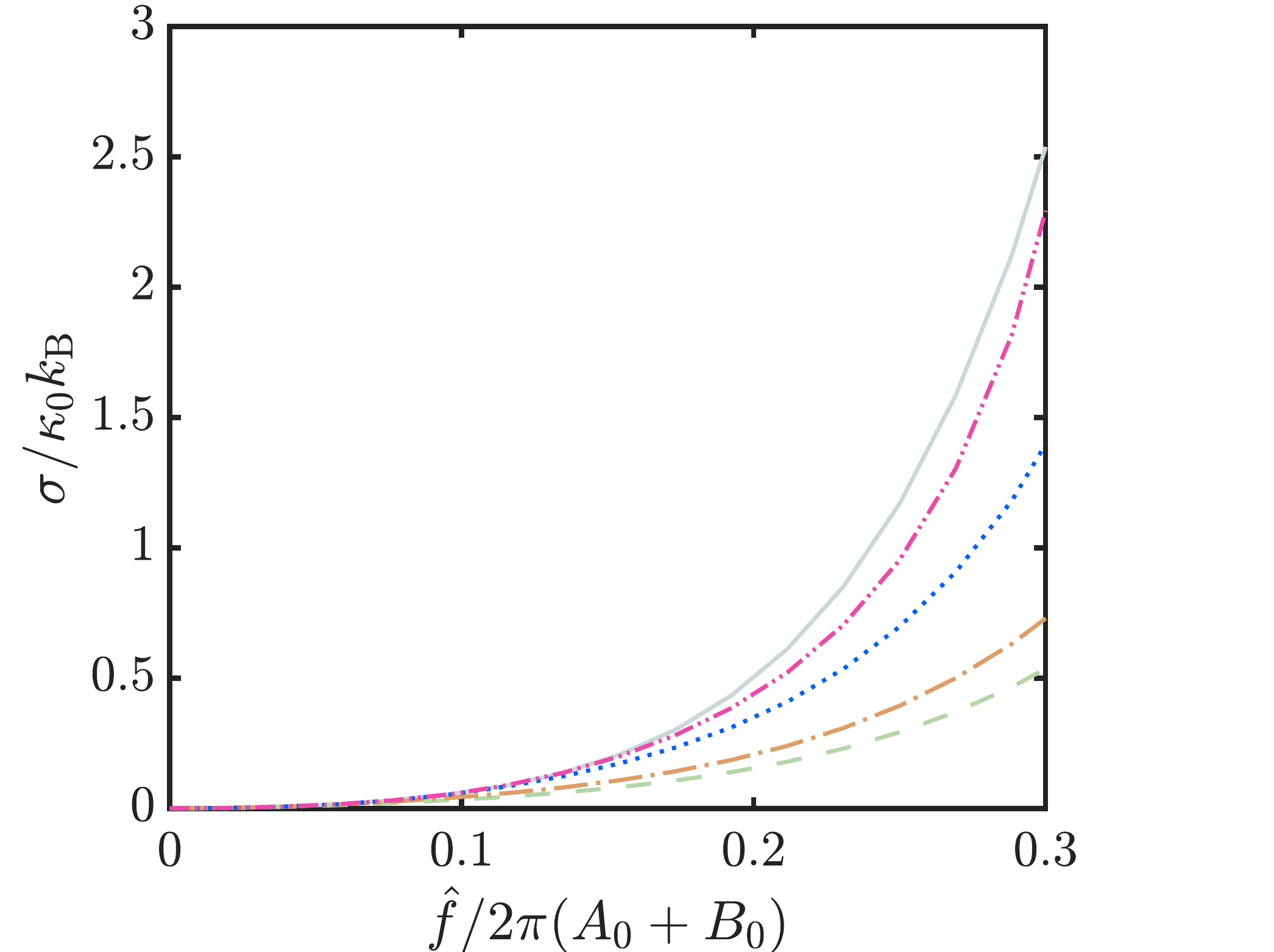}
		\caption{Steady-state entropy production for Eq.\,(\ref{Eq:V:PerTilt}) with periodic part Eq.\,(\ref{Eq:V:Bi}) as function of the tilting force, via: (solid) Eq.\,(\ref{Eq:S:gen}), (dashed) TUR Eq.\,(\ref{Eq:S:TUR}), (dash-dotted) Eq.\,(\ref{Eq:Virial:TUR}) 2-nd order, (dotted) Eq.\,(\ref{Eq:Virial:TUR}) 11-th order, (dash-dot-dot-dashed) Eq.\,(\ref{Eq:Virial:TUR}) 29-th order. Parameters used $A_0=4$, $\nu_A=1$, $B_0=1.2$, $\nu_B=8$.}
		\label{Fig:8}
	\end{center}
\end{figure}

\begin{figure}[htbp]
	\begin{center}
		\includegraphics[width=0.45\textwidth]{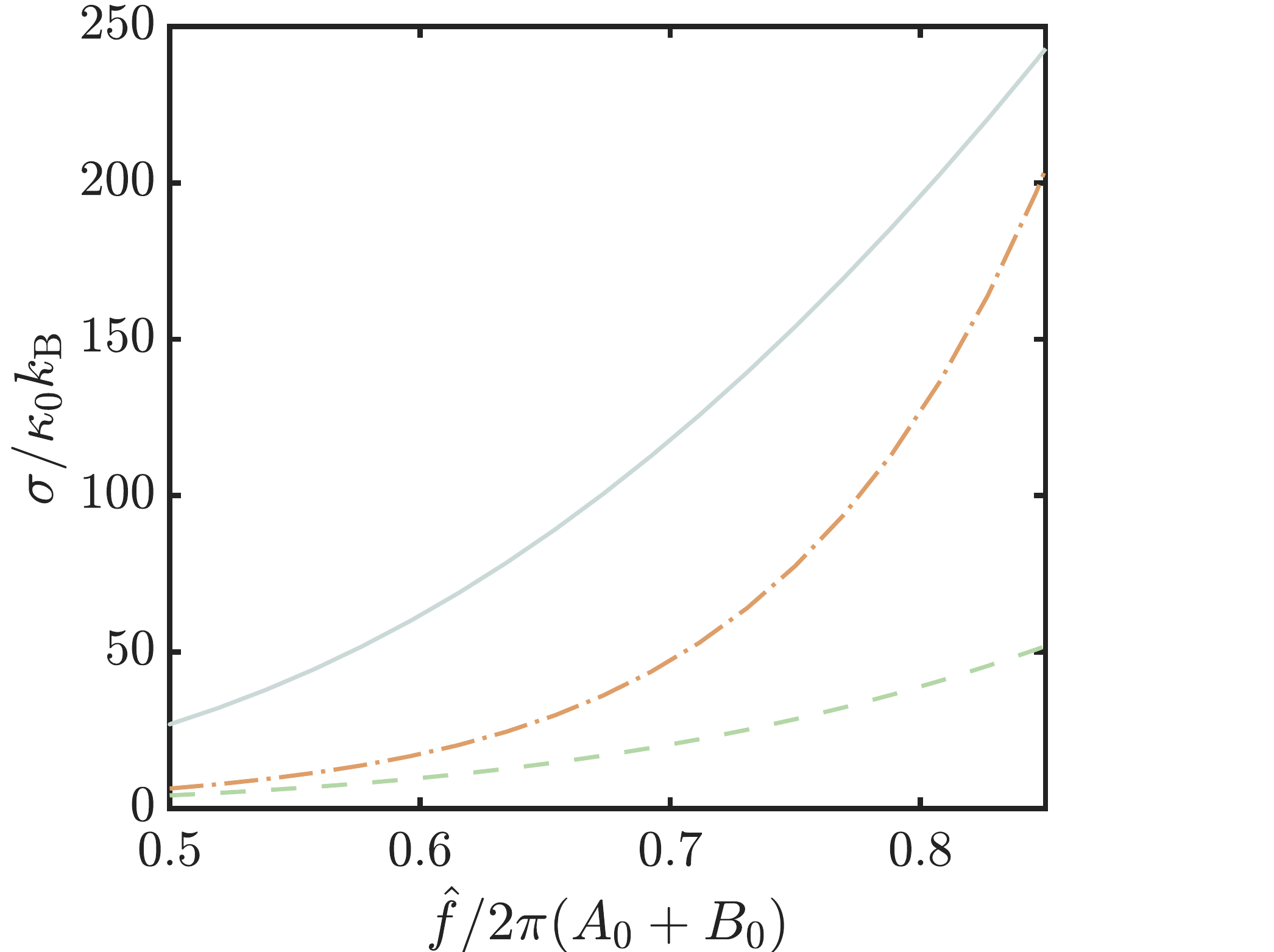}
		\caption{Steady-state entropy production for Eq.\,(\ref{Eq:V:PerTilt}) with periodic part Eq.\,(\ref{Eq:V:Bi}) in the transient regime, via: (solid) Eq.\,(\ref{Eq:S:gen}), (dashed) TUR Eq.\,(\ref{Eq:S:TUR}) and (dash-dotted) Eq.\,(\ref{Eq:Virial:TUR}) 2-nd order. Parameters used $A_0=4$, $\nu_A=1$, $B_0=1.2$, $\nu_B=8$.}
		\label{Fig:9}
	\end{center}
\end{figure}

We now analyse the validity of Eq.\,(\ref{Eq:Virial:TUR}) for a new model potential. We consider Eq.\,(\ref{Eq:V:PerTilt}) periodic part is now
\begin{equation}\label{Eq:V:Bi}
V_0(\hat{x})=k_{\rm B}TA_0\cos{(2\pi\nu_{A}\hat{x})}+k_{\rm B}TB_0\cos{(2\pi\nu_{B}\hat{x})}\,.
\end{equation}  
The different regimes for Eq.\,(\ref{Eq:V:PerTilt}) and periodic part Eq.\,(\ref{Eq:V:Bi}) as we vary the tilting force can be identify via its diffusion\,\cite{Lopez-AlamillaJackChallis2020}, see Figure~\ref{Fig:7}. Again, we compare the TUR lower bound Eq.\,(\ref{Eq:S:TUR}) with the bounds provided by Eq.\,(\ref{Eq:Virial:TUR}) and the actual entropy generation Eq.\,(\ref{Eq:S:gen}) in the tight-binding regime. This is shown in Figure~\ref{Fig:8}, and in Figure~\ref{Fig:9} beyond the tight binding regime. 

\begin{figure}[htbp]
   	\begin{center}
   		\includegraphics[width=0.45\textwidth]{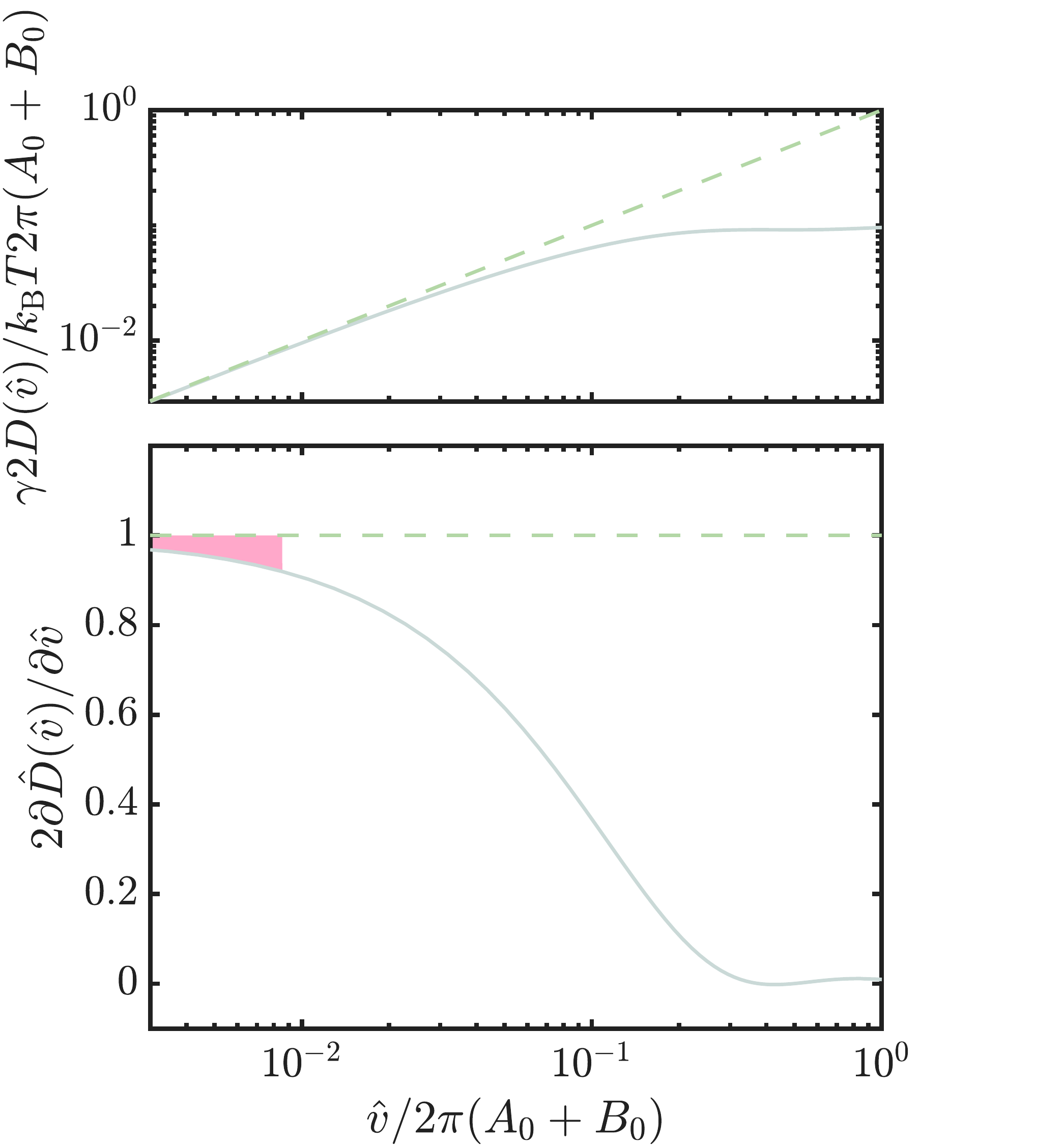}
   		\caption{Evaluation of tight-binding regime validity for Eq.\,(\ref{Eq:V:PerTilt}) with periodic part Eq.\,(\ref{Eq:V:Bi}), showing: (a) Log-log plot of (solid) $2\hat{D}(\hat{v})$ and (dashed) $\hat{v}$. (b) Semi-log plot fo (solid) $2\partial\hat{D}(\hat{v})/\partial\hat{v}$ and (dashed) $\partial\hat{v}/\partial\hat{v}$ for reference. Shaded region shows $2\partial\hat{D}(\hat{v})/\partial\hat{v}\approx1\pm0.1$. Parameters used $A_0=4$, $\nu_A=1$, $B_0=1.2$, $\nu_B=8$.}
   		\label{Fig:10}
   	\end{center}
   \end{figure}

We can use the criteria Eq.\,(\ref{Eq:TB:test}) to quantify the range of the tight-binding regime, for the model periodic potential Eq.\,(\ref{Eq:V:Bi}) and the parameters values of used in Figs.~\ref{Fig:8}, \ref{Fig:9}. This is shown in Figure~\ref{Fig:10}.
\section{Results}

We presented a methodology to approximate the entropy production for Brownian motion in a tilted periodic potential.
The approximation stems from the well known thermodynamic uncertainty relation. By implementing a virial-like expansion of the original TUR we improve the approximation . Equation (\ref{Eq:Virial:TUR}) is particularly useful in the tight-binding regime were both drift and diffusion. In fact, in this regime by increasing the order of the approximation it will converge to the actual value of the entropy generation in the system, see Figures~\ref{Fig:4}, \ref{Fig:8}. It is worth mentioning that the convergence in the tight-binding regime  
of Eq.\,(\ref{Eq:Virial:TUR}) to Eq.\,(\ref{Eq:S:gen}) is quite slow, this is because in this regime $\hat{v}/\hat{D}$ is proportional to the hyperbolic tangent of the applied force. In the transient regime were the stochastic resonance of the diffusion takes place, Eq.\,(\ref{Eq:Virial:TUR}) improves the approximation only for small $N$ values in the expansion, see Figures~\ref{Fig:6}, \ref{Fig:9}. The results presented here for the two model periodic parts considered can be generalized to potentials with more periodicity components. However, the transient regime is larger for more complicated potentials. Thus, it is expected that the force range where those potentials are in the tight-binding regime will be smaller. Ultimately, for any type of periodic potential in the linear regime, Eq.\,(\ref{Eq:Virial:TUR}) will rapidly diverge, except when $N=1$ which in turn returns the original TUR bound of entropy generation.

\bibliography{NJLA-RULC_Virial_TUR_Bibliography}  
\end{document}